\documentclass[%
 aip,
 amsmath,amssymb,
 reprint,%
]{revtex4-1}

\usepackage{graphicx}
\usepackage{dcolumn}
\usepackage{bm}

\usepackage[utf8]{inputenc}
\usepackage[T1]{fontenc}
\usepackage{mathptmx}
\usepackage{etoolbox}

\makeatletter
\def\@email#1#2{%
 \endgroup
 \patchcmd{\titleblock@produce}
  {\frontmatter@RRAPformat}
  {\frontmatter@RRAPformat{\produce@RRAP{*#1\href{mailto:#2}{#2}}}\frontmatter@RRAPformat}
  {}{}
}%
\makeatother
\begin{document}

\preprint{AIP/123-QED}

\title[]{Imaging the photochemical dynamics of cyclobutanone with MeV ultrafast electron diffraction}

\author{Tianyu Wang}
\thanks{These authors contributed equally to this work.}
\affiliation{Key Laboratory for Laser Plasmas (Ministry of Education) and School of Physics and Astronomy,\\Collaborative innovation center for IFSA (CICIFSA),\\Shanghai Jiao Tong University, Shanghai 200240, China.}
\affiliation{Zhangjiang Institute for Advanced Study, Shanghai Jiao Tong University, Shanghai 201210, China.}

\author{Hui Jiang}
\thanks{These authors contributed equally to this work.}
\affiliation{Key Laboratory for Laser Plasmas (Ministry of Education) and School of Physics and Astronomy,\\Collaborative innovation center for IFSA (CICIFSA),\\Shanghai Jiao Tong University, Shanghai 200240, China.}
\affiliation{Zhangjiang Institute for Advanced Study, Shanghai Jiao Tong University, Shanghai 201210, China.}
\affiliation{Tsung-Dao Lee Institute, Shanghai Jiao Tong University, Shanghai 201210, China.}

\author{Cheng Jin}
\affiliation{Key Laboratory for Laser Plasmas (Ministry of Education) and School of Physics and Astronomy,\\Collaborative innovation center for IFSA (CICIFSA),\\Shanghai Jiao Tong University, Shanghai 200240, China.}
\affiliation{Zhangjiang Institute for Advanced Study, Shanghai Jiao Tong University, Shanghai 201210, China.}

\author{Xiao Zou}
\affiliation{Key Laboratory for Laser Plasmas (Ministry of Education) and School of Physics and Astronomy,\\Collaborative innovation center for IFSA (CICIFSA),\\Shanghai Jiao Tong University, Shanghai 200240, China.}
\affiliation{Zhangjiang Institute for Advanced Study, Shanghai Jiao Tong University, Shanghai 201210, China.}

\author{Pengfei Zhu}
\affiliation{Key Laboratory for Laser Plasmas (Ministry of Education) and School of Physics and Astronomy,\\Collaborative innovation center for IFSA (CICIFSA),\\Shanghai Jiao Tong University, Shanghai 200240, China.}
\affiliation{Tsung-Dao Lee Institute, Shanghai Jiao Tong University, Shanghai 201210, China.}

\author{Tao Jiang}
\affiliation{Key Laboratory for Laser Plasmas (Ministry of Education) and School of Physics and Astronomy,\\Collaborative innovation center for IFSA (CICIFSA),\\Shanghai Jiao Tong University, Shanghai 200240, China.}
\affiliation{Zhangjiang Institute for Advanced Study, Shanghai Jiao Tong University, Shanghai 201210, China.}

\author{Feng He}
\affiliation{Key Laboratory for Laser Plasmas (Ministry of Education) and School of Physics and Astronomy,\\Collaborative innovation center for IFSA (CICIFSA),\\Shanghai Jiao Tong University, Shanghai 200240, China.}

\author{Dao Xiang}
\thanks{Corresponding author: \href{mailto:dxiang@sjtu.edu.cn}{dxiang@sjtu.edu.cn}}
\email{dxiang@sjtu.edu.cn}
\affiliation{Key Laboratory for Laser Plasmas (Ministry of Education) and School of Physics and Astronomy,\\Collaborative innovation center for IFSA (CICIFSA),\\Shanghai Jiao Tong University, Shanghai 200240, China.}
\affiliation{Zhangjiang Institute for Advanced Study, Shanghai Jiao Tong University, Shanghai 201210, China.}
\affiliation{Tsung-Dao Lee Institute, Shanghai Jiao Tong University, Shanghai 201210, China.}

\date{\today}

\begin{abstract}
We study the photoinduced chemical dynamics of cyclobutanone upon excitation at 200 nm to the 3s Rydberg state using MeV ultrafast electron diffraction (UED). We observe both the elastic scattering signal, which contains information about the structural dynamics, and the inelastic scattering signal, which encodes information about the electronic state. Our results suggest a sub-picosecond timescale for the photodissociation dynamics, and an excited state lifetime of about 230 femtoseconds. The dissociation is found to be dominated by the C$_3$ channel where cyclopropane and CO are produced. The branching ratio of the C$_3$ channel to the C$_2$ channel where ethene and ketene are produced, is estimated to be approximately 5:3. Our data suggest that the C$_3$ and C$_2$ channels account for approximately 80\% of the photoproducts, with the remaining 20\% exhibiting ring-opened structures. It is found that the timescale associated with the dissociation process in the C$_2$ channel is shorter compared to that in the C$_3$ channel. Leveraging the enhanced temporal resolution of MeV UED, our results provide a real-time mapping of the nuclear wavepacket dynamics, capturing the complete photochemical dynamics from S$_2$ minimum through the S$_1$/S$_0$ conical intersection, and finally to the dissociation. Our experimental results provide new insights into the Norrish Type I reaction and can be used to benchmark non-adiabatic dynamics simulations.
\end{abstract}

\maketitle

\section{\label{sec:level1}Introduction}
Fragmentation of a chemical species upon absorption of light is one of the most important processes in photophysics, photochemistry, and photobiology. A classic example is the Norrish Type I reaction~\cite{NORRISH1937}, where photoexcitation of an aliphatic ketone leads to cleavage of the carbon-carbon bond between the $\alpha$-carbon and the carbonyl group. Cyclobutanone (C$_4$H$_6$O), with only one four-carbon ring, has been the subject of intense study for many decades to understand the strain effect in photochemistry and to gain insight into the photodissociation dynamics of the carbonyl compounds~\cite{Benson1942,Denschlag1967,Campbell1967,TURRO1967,Lee1969,whitlock1971,Hemminger1972,Hemminger1973,Tang1976,Trentelman1990,diau2001,Kuhlman2012,kuhlman2013,liu2016}. Recently, cyclobutanone was chosen as a model system for a prediction challenge in which fifteen groups reported their independent predictions of the key dynamics and experimental observables (e.g. time-resolved diffraction pattern with electron scattering) of this molecule upon photoexcitation at 200 nm ~\cite{eng2024,Vindel-Zandbergen2024,miao2024,peng2024,janos2024,hutton2024,makhov2024,suchan2024,jaiswal2024,lawrence2024,mukherjee2024,martin_santa2024,hait2024,miller2024,bennett2024}. In this study, we present our experimental results on the photoinduced ultrafast dynamics of cyclobutanone, obtained through MeV UED, a scattering technique with direct sensitivity to structural dynamics for samples in solid, liquid and gas phases~\cite{zong2018ultrafast,Yang2018,Sie2019,Wolf2019,Yang2020,Yang2021,Champenois2021,Lin2021,Duan2021,Cheng2022,Ma2022,Wu2022,lu2022photoinduced,Xu2023,Champenois2023,Cheng2024,Heo2024}. The experiment is conducted under conditions analogous to those employed in the simulations, thereby facilitating direct comparisons between theoretical predictions and experimental observations.

Cyclobutanone has a weak absorption band from 330 nm to 240 nm and a moderate absorption band from 204 nm to 180 nm~\cite{whitlock1971}. The weak absorption is related to the n-$\pi^*$ transition and the molecule is excited to the lowest singlet excited state (S$_1$) \cite{Hemminger1973}. The moderate absorption band is due to the n-3s Rydberg excitation that promotes the molecules to the second excited singlet state (S$_2$)~\cite{whitlock1971}. Extensive measurements have been made for cyclobutanone upon excitation to S$_1$ and the molecule is found to undergo various photochemical reactions including two primary dissociation channels: the C$_2$ channel where the photoproducts are thene and ketene, and the C$_3$ channel where cyclopropane and CO are formed~\cite{Benson1942,Denschlag1967,TURRO1967}. The branching ratio of the C$_3$ and C$_2$ products (C3:C2) ranges from 0.4 to 7 and is reported to be wavelength dependent~\cite{Campbell1967,Lee1969,Hemminger1972,Tang1976,Trentelman1990}. In spite of extensive studies of cyclobutanone excited to S$_1$, only a few measurements have been made when the molecule has been excited to S$_2$. For example, in the 193 nm photolysis of cyclobutanone~\cite{Trentelman1990}, rotationally hot CO was found to account for 85\% of the total CO yield. The high-rotational-temperature major product is assigned to the C$_3$ channel, while the minor product is assigned to the C$_2$ channel, where the ketene dissociates further to produce low-rotational-temperature CO. Since it is difficult to estimate how much of the ketene dissociates to give methylene and CO, the value of 85:15 may be considered as the upper limit of the branching ratio of C$_3$:C$_2$ at 193 nm excitation. Regarding the time scale of the dynamics in S$_2$, a decay constant of 0.74 ps was observed by time-resolved mass spectrometry and time-resolved photoelectron spectroscopy~\cite{Kuhlman2012}. This time constant was inferred to be related to the internal conversion to the S$_1$ state.  
\begin{figure}
    \centering
    \includegraphics[width=1\linewidth]{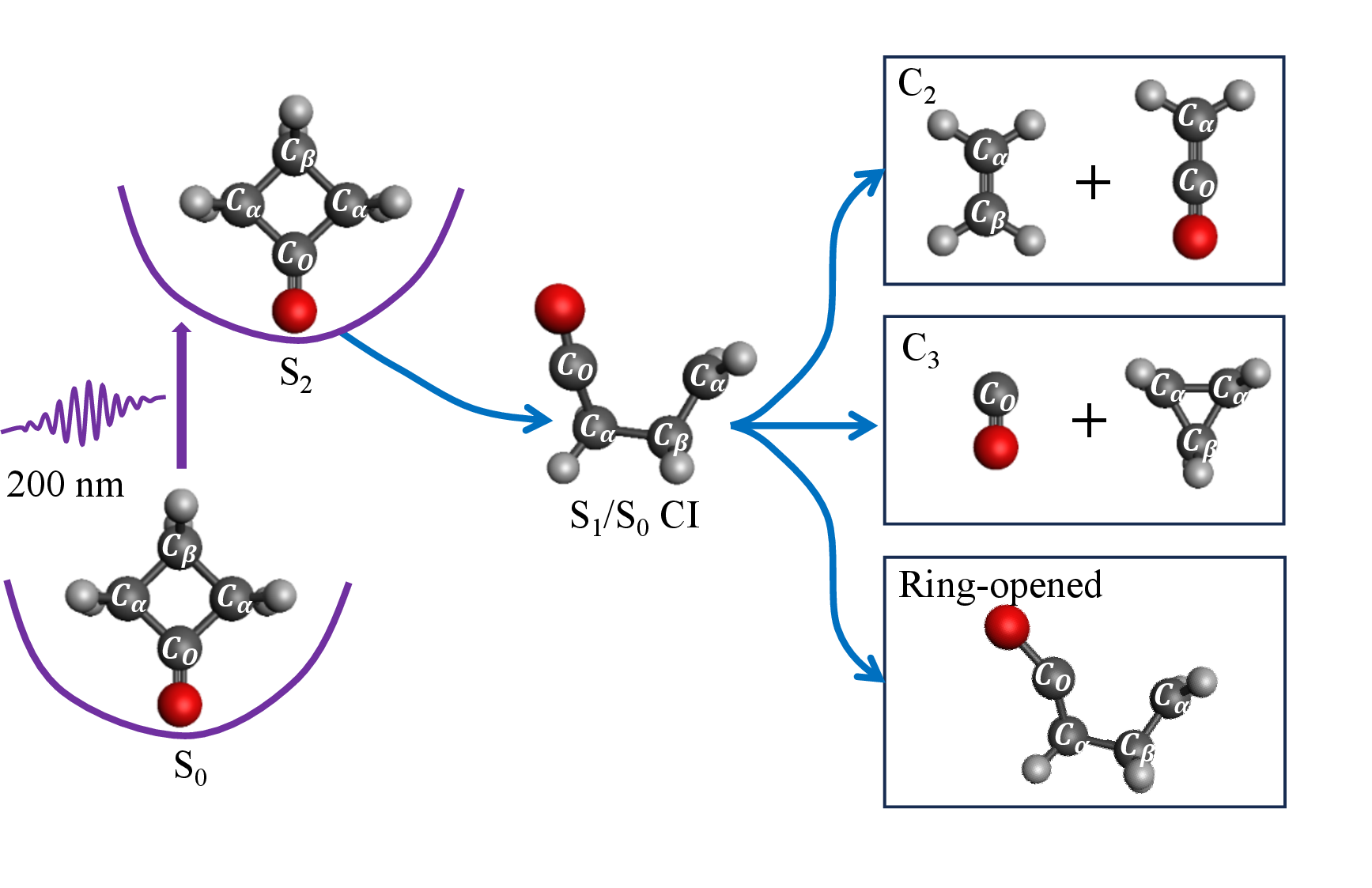} 
    \caption{\label{f1}The schematic reaction pathway of cyclobutanone.}
\end{figure}

In light of recent comprehensive theoretical investigations, the streamlined chemical dynamics of cyclobutanone under photoexcitation at 200 nm are depicted schematically in Fig.~\ref{f1}. Following deep UV photoexcitation from the ground state (S$_0$) to the 3s Rydberg state (S$_2$), ring puckering and CO stretching are induced which serve as the driving force for the subsequent ring-opening and dissociation. A nonplanar-to-planar transition occurs in the molecule's structure as it relaxes from the Franck-Condon region to the S$_2$ minimum. Given the nature of the Rydberg state as a bound state, direct photodissociation from S$_2$ is unlikely and the changes in the molecular geometry should be minor. The molecule is expected to further relax to the S$_1$ state and then proceed along the S$_1$ potential energy surface to reach the S$_1$/S$_0$ conical intersection (CI). At the CI, an open structure will be formed, and the large excess energies will be converted into the C-C stretching modes that further trigger the subsequent dissociation. The wave packet is expected to return to the ground state, producing thene and ketene in the C$_2$ channel and cyclopropane and CO in the C$_3$ channel. It is also possible that the molecule may relax to the ground state with ring-opened cyclobutanone, similar to the electrocyclic ring-opening reaction of 1,3-cyclohexadiene~\cite{Attar2017,Wolf2019}. The primary objective of the prediction challenge is to ascertain the lifetime of the excited state, the structural dynamics during photodissociation, and the branching ratio of the various photoproducts.

\section{Experimental details and static measurements}

\begin{figure*}
    \centering
    \includegraphics[width=1\linewidth]{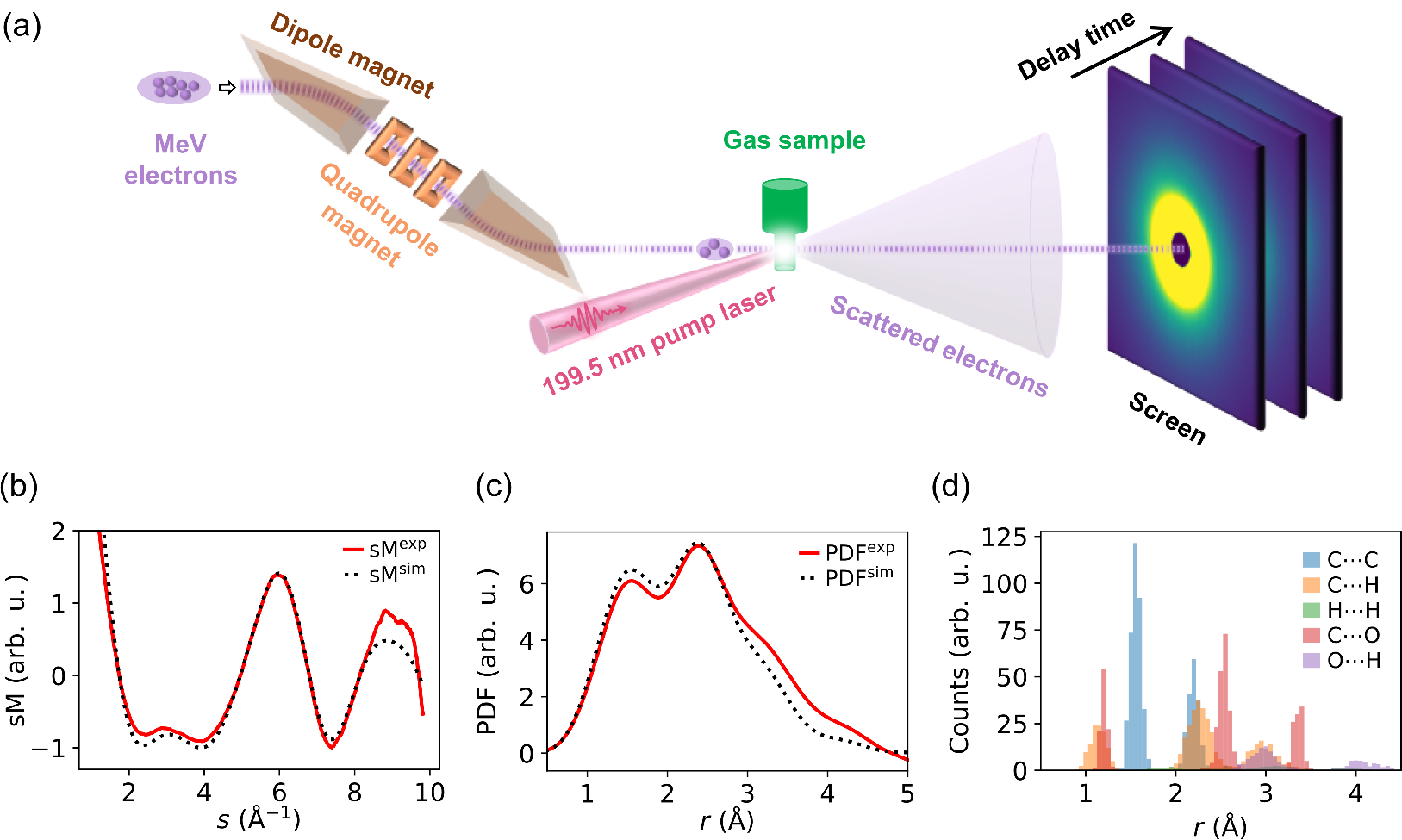} 
    \caption{\label{f2} The experimental setup and the stationary measurement results. (a) Schematic of the MeV UED instrument. The electron beam is compressed by a double bend achromat (DBA) system, which consists of two dipole magnets and three quadrupole magnets. This configuration serves to reduce both the electron pulse width and the arrival time jitter. Thereafter, the electron beam intersects with the focused 199.5-nm laser pulse at the gas sample, which is delivered via a flow cell. The scattered electrons are measured with a phosphor screen, which is imaged to an EMCCD. (b) Comparison between experimental (red solid) and simulated (black dotted) static modified molecular diffraction intensity (sM). (c) Comparison between experimental (red solid) and simulated (black dotted) static pair distribution function (PDF). (d) The simulated static pair distance distribution scaled by the product of atomic scattering factors.}
\end{figure*}

In order to be able to compare the data directly with the simulation results, our experiment was performed under similar conditions as the simulations. The measurement was performed using the MeV UED instrument~\cite{Qi2020} at Shanghai Jiao Tong University (SJTU) and the experimental setup is sketched in Fig.~\ref{f2}(a). Briefly, a UV laser pump pulse (199.5~nm, about 120~fs full width at half maximum, FWHM) is used to excite the cyclobutanone molecules to the 3s Rydberg state in the gas phase, initiating the Norrish Type I reaction. An ultrashort electron beam (3 MeV kinetic energy), generated using a photocathode radio-frequency gun and subsequently compressed in a double-bend achromat (DBA) lens, is then used to probe the dynamics. The DBA compressor functions to reduce both the electron pulse width (about 40 fs) and the timing jitter (about 20 fs), thereby ensuring an overall temporal resolution of approximately 130 fs (FWHM) in this measurement. The cyclobutanone sample is procured from Adamas without undergoing additional purification. The gas is delivered with a flow cell measuring 2~mm in length and 0.7 mm in diameter. The gas cell is heated to $\sim$85~$^\circ$C to mitigate the occurrence of sample condensation. The electron beam size at the gas cell is measured to be about 150~$\mu$m (FWHM). The UV laser spot size at the gas cell is approximately 250~$\mu$m (FWHM) and the pulse energy is about 10~$\mu$J to avoid multiphoton process and ionization. The diffraction pattern is measured with a phosphor screen imaged to an electron-multiplying charge-coupled device (EMCCD). The phosphor screen contains a 3-millimeter diameter hole in the center, allowing for the transmission of unscattered electrons that would otherwise saturate the detector. Consequently, the data within the range of $s<0.8~\text{\AA}^{-1}$ are not measured. The electron beam repetition rate is 400 Hz and each scattering pattern at a designated time delay is accumulated over a period of 5 seconds with 2000 electron pulses, each with an approximate charge of 20~fC. The diffraction pattern is measured with a time step of 53~fs in the time window from -0.6~ps to 1.2~ps. The full data sets include 250 such scans and the total integration time at each time delay is about 20 minutes.

As the scattering signal diminishes rapidly with increasing momentum transfer ($s$) of the scattered electron, the measured static scattering signal of ground-state cyclobutanone is represented by the modified molecular diffraction intensity sM$(s)$~\cite{Centurion2022}, as shown in Fig.~\ref{f2}(b). The measurement is in excellent agreement with the simulation based on the sampled geometries. The real-space pair distribution function (PDF), which represents the probability of finding an atomic pair at distance $r$, is obtained by performing a Fourier-sine transform on the static sM,
\begin{equation}
    {\rm{PDF}}(r)=r\int_{s_{min}}^{s_{max}} {\rm{sM}}(s)\sin(sr)e^{-ks^2}ds,
    \label{cal_PDF}
\end{equation}
where the function $e^{-ks^2}$ is used to prevent non-physical signals caused by truncation in the momentum space. This function also serves to mitigate the effect of increased noise in the high-$s$ region. In this study, $k$ is set to 0.03~$\text{\AA}^2$ and $s_{max}$ is taken to be $10~\text{\AA}^{-1}$. 

As mentioned previously, the signals at $s<0.8$~$\text{\AA}^{-1}$ are not measured due to the central hole of the phosphor screen. To circumvent the occurrence of non-physical outcomes stemming from the truncation of the low-$s$ region, scaled simulation results are used to supplement the data within the range of $0<s<0.8$~$\text{\AA}^{-1}$. The corresponding static PDF is shown in Fig.~\ref{f2}(c) where two broad peaks, centered at approximately 1.6~$\text{\AA}$ and 2.4~$\text{\AA}$, are clearly visible. A comparison to the simulated statistical atomic pair distribution (Fig.~\ref{f2}(d)) of the cyclobutanone molecule suggests that the first peak in the measured PDF includes contributions from the nearest C-C, nearest C-H, and nearest C-O atomic pairs. The second peak is associated with the diagonal C-C, the second nearest C-H, and the second nearest C-O atomic pairs. Given the proportionality of the PDF's intensity to the scattering cross sections of atomic pairs and the number of atomic pairs, the C-H atomic pairs yield a substantial intensity in the PDF, despite H atom's comparatively small cross section. It is noteworthy that the diffraction limit results in a width of approximately $0.6~\text{\AA}$ for the PDF peaks. Consequently, the specific atomic pairs with small interatomic distance differences cannot be distinguished in the PDF and only two broad peaks are observed.

\section{Time-resolved measurement of diffraction difference}

\begin{figure*}
    \centering
    \includegraphics[width=0.9\linewidth]{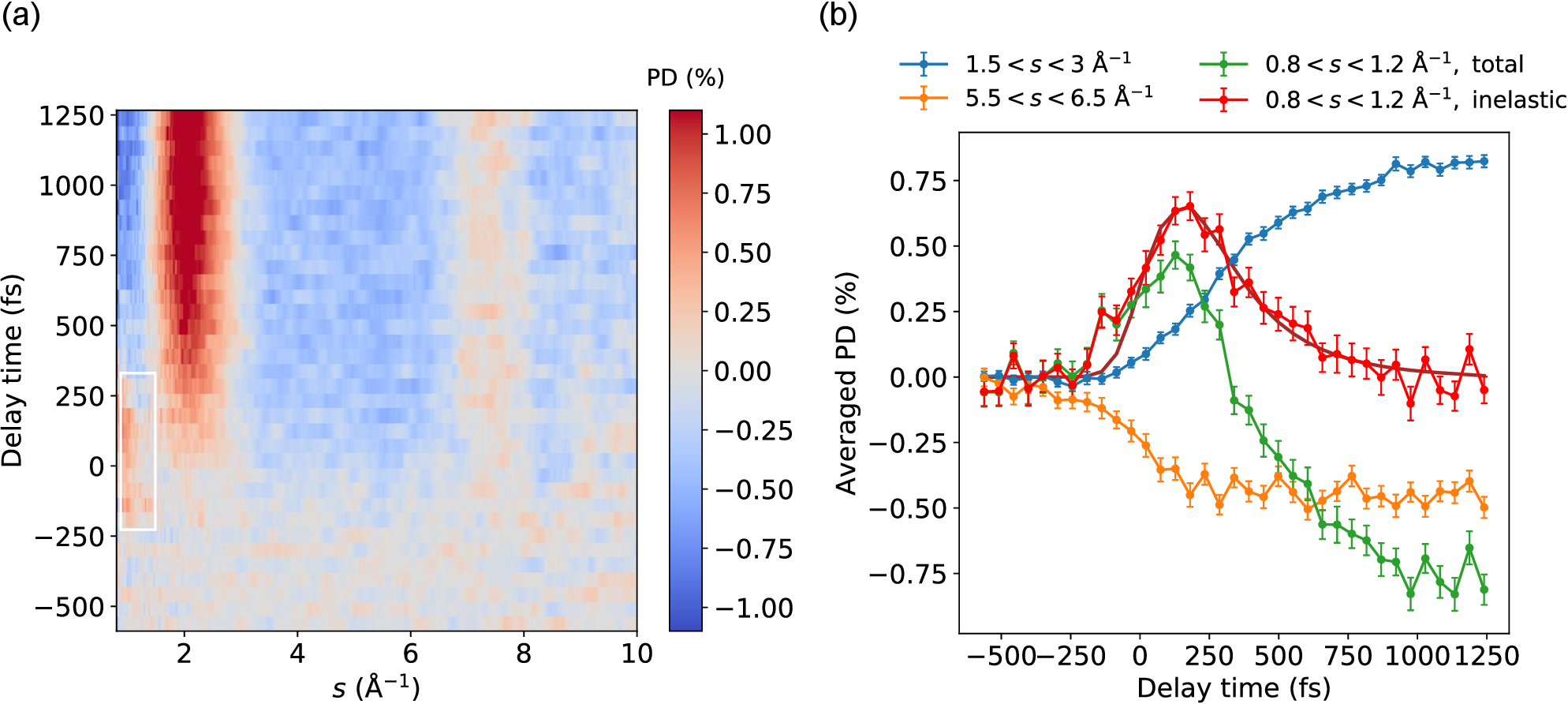} 
    \caption{\label{f3} Time-resolved measurements in momentum space. (a) The experimental percentage difference (PD) as a function of the delay time. The small-angle inelastic scattering signals are indicated by the white rectangle. (b) The averaged PD in various $s$-ranges, i.e. $1.5<s<3$~$\text{\AA}^{-1}$ (blue), $5.5<s<6.5$~$\text{\AA}^{-1}$ (orange), the small angle regime $0.8<s<1.2$~$\text{\AA}^{-1}$ (green).  The red line signifies the corrected inelastic scattering signal. The rising edge of the corrected inelastic scattering signal is fitted with an error function to determine the time zero, while the falling edge is fitted by the convolution of the instrumental response function and an exponential decay function. The brown line shows the fitting results, and the fitted decay constant is $230\pm40$ fs. The error bars represent one standard deviation calculated through bootstrap.}
\end{figure*}

The measured time-resolved percent-difference (PD) diffraction signal is shown in Fig.~\ref{f3}(a). PD is defined as,
\begin{equation}
    {\rm{PD}}(s,t)=\frac{I(s,t)-I(s,t<0)}{I(s,t<0)}.
        \label{cal_PD}
\end{equation}
where $I(s,t)$ is the measured diffraction signal at momentum transfer $s$ and time delay $t$, and  $I(s,t<0)$ is the signal when the electron beam arrives at the sample earlier than the pump laser. 
This diffraction-difference signal effectively eliminates the large contributions from unexcited molecules and allows the detection of small signals related to structural changes in the excited molecules, which typically represent only a small fraction of the total molecules. It also effectively removes the background from the dark current and the pump laser. The measured PD signal is characterized by positive and negative changes as a function of time delay and momentum transfer. Specifically, enhanced bands are observed for the regions $1.5<s<3~\text{\AA}^{-1}$ and  $6.5<s<8~\text{\AA}^{-1}$, while bleached bands appear in the regions $s<1.5~\text{\AA}^{-1}$, $3<s<6.5~\text{\AA}^{-1}$ and $8<s<10~\text{\AA}^{-1}$. The PD signal shows negligible changes after 1~ps, indicating ultrafast dynamics of the photochemical reaction. The overall characteristics of the measured PD are consistent with simulations predicting sub-picosecond dissociation~\cite{eng2024,Vindel-Zandbergen2024,miao2024,peng2024,janos2024,hutton2024,makhov2024,suchan2024,jaiswal2024,lawrence2024,mukherjee2024,martin_santa2024}.  

It should be noted that an enhanced signal is observed in the very low momentum transfer region ($s<1.2~\text{\AA}^{-1}$, as indicated by the white rectangle) immediately after photoexcitation. This anomalous signal is transient, lasting only a brief period, in stark contrast to the signal observed in the $1.2<s<10~\text{\AA}^{-1}$ region, which persists for the entire duration of the measurement. This signal is likely due to inelastic scattering associated with electronic state changes previously observed in other systems, including pyridine~\cite{Yang2020} and ammonia~\cite{Champenois2023}. Given that the independent atom model (IAM) does not account for electron-electron correlation~\cite{Yang2020}, this inelastic scattering signal is absent in all theoretical predictions~\cite{eng2024,Vindel-Zandbergen2024,miao2024,peng2024,janos2024,hutton2024,makhov2024,suchan2024,jaiswal2024,lawrence2024,mukherjee2024,martin_santa2024,hait2024,miller2024,bennett2024} where the PD is calculated from the simulated trajectories using the IAM. 

The time dependence of the PD signals in three representative momentum transfer regions is shown in Fig.~\ref{f3}(b). The PD signal at $0.8<s<1.2~\text{\AA}^{-1}$ (green line in Fig.~\ref{f3}(b)) is attributed to both inelastic and elastic scattering. Consequently, it manifests a positive signal at the early stage, which is predominantly influenced by inelastic scattering, and a negative signal in the subsequent stage, which is dominated by elastic scattering associated with dissociation. Given the observation of a negative signal at $s<1.2~\text{\AA}^{-1}$ that changes in synchrony with the positive signal at $1.5<s<3.0~\text{\AA}^{-1}$ as demonstrated in the simulations~\cite{eng2024,miao2024,peng2024,janos2024,hutton2024,makhov2024}, it is possible to obtain the elastic scattering signal at $0.8<s<1.2~\text{\AA}^{-1}$ in the early stage with extrapolations derived from experimental data within the range of $1.5<s<3.0~\text{\AA}^{-1}$. This methodology enables the extraction of a corrected inelastic scattering signal (illustrated as the red line in Fig.~\ref{f3}(b)) through the subtraction of contributions from elastic scattering. The rising edge of the signal is modeled with an error function, with the center of the function designated as the time zero of the measurement. The falling edge of the signal is modeled with a convolution of the instrument response function and a Gaussian function, yielding a decay constant of approximately 230~fs. This time constant can be interpreted as the lifetime of the excited state; however, it is difficult to separate the S$_1$ state from S$_2$ state with the available data. Further theoretical investigation is necessary to accurately quantify the individual contributions of S$_2$ and S$_1$ states to the inelastic signal, a topic that extends beyond the scope of the present study. 


For the elastic signal, a discernible delay is evident in the low-$s$ signal ($1.5<s<3.0~\text{\AA}^{-1}$, blue line in Fig.~\ref{f3}(b)) in comparison to the high-$s$ signal ($5.5<s<6.5~\text{\AA}^{-1}$, orange line in Fig.~\ref{f3}(b)). This can be understood in terms of the inverse relationship between momentum space and real space, i.e. the small distance changes in real space result in signal changes in the high momentum transfer region and vice versa. Notably, at the earliest stage preceding dissociation, significant changes are predominantly observed in the high-$s$ region, since the signal at high momentum transfer is sensitive to structural changes at short distances. In contrast, the signal in the low-$s$ region gains significant value only after the dissociation starts, because the signal at small momentum transfer is sensitive to structural changes at large distances. The time at which significant changes become evident in the high-$s$ signal is found to coincide with that of the inelastic signal within our measurement uncertainty, indicating a rapid and subtle change of the molecular structure immediately following photoexcitation.

\section{Real-space inversion of PDF}

\begin{figure*}
    \centering
    \includegraphics[width=0.9\linewidth]{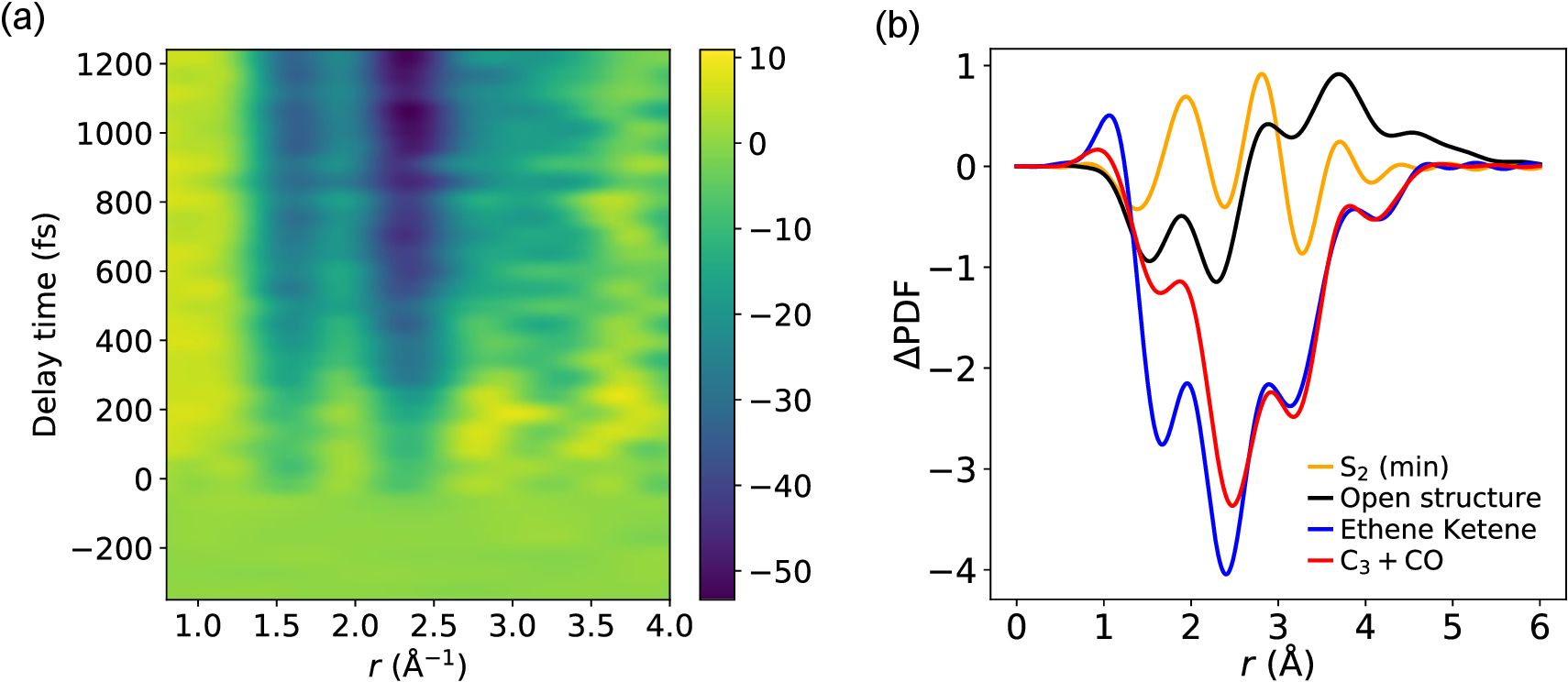} 
    \caption{\label{f4} Time-resolved measurements in real space. (a) Experimental $\Delta$PDF as a function of delay time. (b) The simulated $\Delta$PDF for a variety of molecule geometries. The orange, black, blue and red lines represent the results for S$_2$ minimum, open structure at S$_1$/S$_0$ CI, ethene ketene and C$_3$+CO, respectively.  }
\end{figure*}

\begin{figure*}
    \centering
    \includegraphics[width=0.9\linewidth]{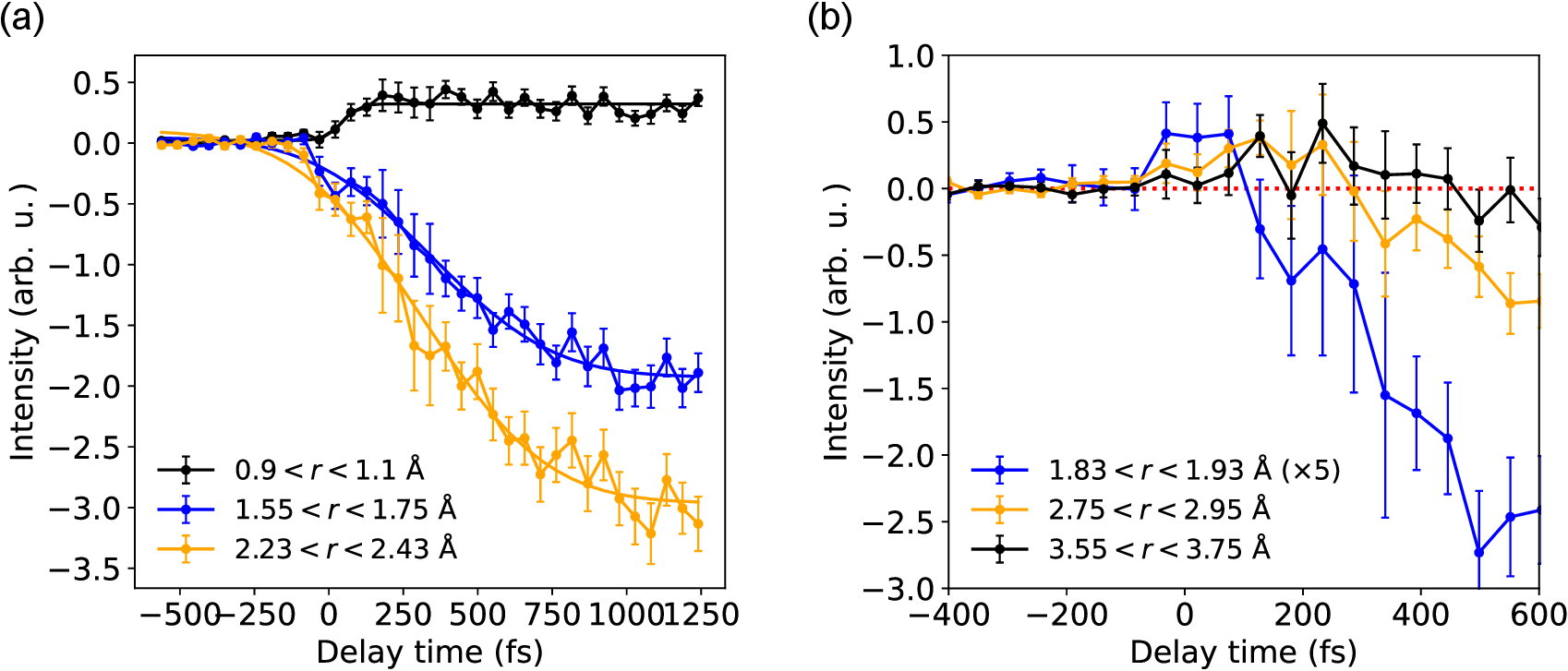} 
    \caption{\label{f5} Temporal evolution of the intensity of the measured $\Delta$PDF. (a) The black, blue and orange lines represent the intensity of $\Delta$PDF in the range of $0.9<r<1.1~\text{\AA}$, $1.55<r<1.75~\text{\AA}$ and $2.23<r<2.43~\text{\AA}$, respectively. (b) The blue, orange and black lines represent the intensity of $\Delta$PDF in the range of $1.83<r<1.93~\text{\AA}$, $2.75<r<2.95~\text{\AA}$ and $3.55<r<3.75~\text{\AA}$, respectively. The error bars represent one standard deviation calculated through bootstrap.}
\end{figure*}

The large momentum transfer range accessible in UED measurement enables precise inversion to obtain the PDF in real space using the diffraction data. To mitigate the impact of artifacts, the missing information in the range of $s<0.8~\text{\AA}^{-1}$ is extrapolated from the time-resolved diffraction data where the signal in the range of $0.8<s<1.2~\text{\AA}^{-1}$ is corrected using the aforementioned methodology. Figure~\ref{f4}(a) shows the measured  difference PDF ($\Delta$PDF) obtained from a Fourier-sine transform of $\Delta$sM. In order to achieve a more profound comprehension of the characteristics manifested by the measured $\Delta$PDF, the calculated  $\Delta$PDF for various molecule geometries and photoproducts is presented in Fig.~\ref{f4}(b).

The most prominent features in the measured $\Delta$PDF are an enhanced band at $r\approx1.0~\text{\AA}$ and two bleached bands at $r\approx1.6~\text{\AA}$ and $r\approx2.3~\text{\AA}$. These features agree well with the $\Delta$PDF associated with the dissociation in C$_2$ and C$_3$ channels (blue and red lines in Fig.~\ref{f4}(b)). For the C$_3$ channel, the cleavage of the two C$_\text{O}$-C$_\alpha$ bonds results in a reduced number of atomic pairs at the nearest C-C distance, which accounts for the negative signal at $r\approx1.6~\text{\AA}$. The intensity at $r\approx2.3~\text{\AA}$ is also reduced due to the loss of the O-C$_\alpha$ pair and the diagonal C-C pairs. While analogous changes are also predicted in the C$_2$ channel, the production of CH$_2$CO in the C$_2$ channel with reduced O-C$_\text{O}$ distance results in a positive signal at $r\approx1.0~\text{\AA}$. This distinctive feature is exclusively associated with the C$_2$ channel and may be utilized to monitor its dynamics~\cite{janos2024}. The time dependence of the $\Delta$PDF in these three interatomic distances is shown in Fig.~\ref{f5}(a). The intensity at 1.0~\text{\AA} is nearly saturated after 250 fs while that at 1.6~\text{\AA} and 2.3~\text{\AA} continues to decrease. This observation indicates that the timescale associated with the dissociation process in the C$_2$ channel is shorter compared to that in the C$_3$ channel, which is consistent with the theoretical predictions~\cite{janos2024,lawrence2024}.

\begin{figure*}
    \centering
    \includegraphics[width=1\linewidth]{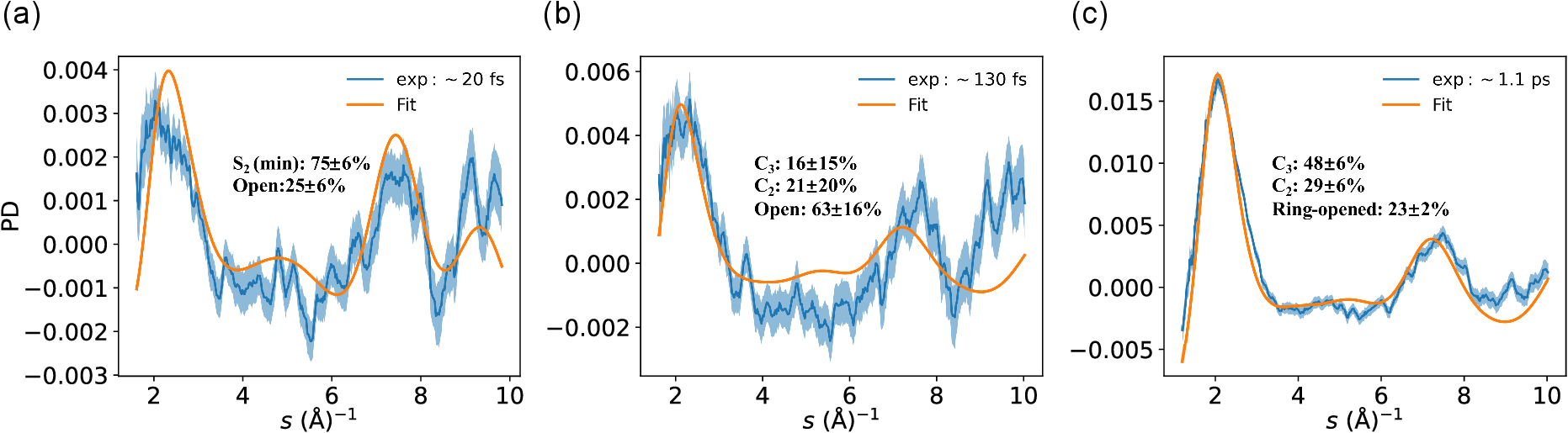} 
    \caption{\label{f6} The fitting of the branching ratio of the photoproducts based on the measured PD at various time delays. The fitting results are determined using a global optimization approach with basis functions of five photoproducts: S$_2$ (min), open structure at S$_1$/S$_0$ CI, C$_2$ channel, C$_3$ channel and the ring-opened structure. The measured PDs are averaged over a $\sim$100~fs time window. (a) The fitting results indicate that the open structure accounts for $25\pm6\%$ and S$_2$ (min) accounts for $75\pm6\%$ at $t\approx$20~fs. (b) The fitting results show that the open structure accounts for $63\pm16\%$, C$_3$ accounts for $16\pm15\%$ and C$_2$ accounts for $21\pm20\%$ at $t\approx$130~fs. (c) The fitting results show that the ring-opened structure accounts for $23\pm2\%$, C$_3$ accounts for $48\pm6\%$ and C$_2$ accounts for $29\pm6\%$ at $t\approx$1.1~ps. The error bars represent one standard deviation calculated through bootstrap.}
\end{figure*}

In addition to the aforementioned dissociation-related three bands that persist for the entire duration of the measurement, transient positive features that only last for a brief period, are also observed. Specifically, two positive signals at approximately 1.9~\text{\AA} and 2.8~\text{\AA} manifest around time zero. These features are consistent with the  $\Delta$PDF associated with the geometry of the molecule at S$_2$ minimum (orange line in Fig.~\ref{f4}(b)). Upon photoexcitation from S$_0$ to S$_2$, the relaxation from the Franck-Condon region to S$_2$ minimum is expected to involve an increase in the  C$_\text{O}$-C$_\alpha$ distance due to planarization, which also results in an increase in the O-C$_\alpha$ distance~\cite{janos2024,peng2024}. The lengthening of the interatomic distance would result in negative signals at the initial interatomic distances (e.g. 1.6~\text{\AA} and 2.5~\text{\AA}) and positive signals at slightly larger distances ((e.g. 1.9~\text{\AA} and 2.8~\text{\AA}). The resulting modulation in the $\Delta$PDF is in good agreement with the measurement. It is noteworthy that a minor alteration in the molecular geometry may result in a substantial modulation in $\Delta$PDF, a unique characteristic that allows resolving subtle structural changes in UED\cite{ihee2001direct}. For instance, a transient contraction of a hydrogen bond by 0.004~\text{\AA} has been observed to lead to an \text{\AA}-scale modulation of the $\Delta$PDF\cite{Yang2021}. 

The time dependence of the $\Delta$PDF at approximately 1.9~\text{\AA}, 2.8~\text{\AA} and 3.7~\text{\AA} is shown in Fig.~\ref{f5}(b). The positive signal at 1.9~\text{\AA} is particularly brief, with a prompt increase following photoexcitation and a subsequent decline to zero at around 100 fs. The signal at 2.8~\text{\AA} lasts approximately twice as long as that at 1.9~\text{\AA}. This behavior is likely associated with the subsequent structural change in the molecule structure as it crosses the S$_1$/S$_0$ CI. As shown in Fig.~\ref{f4}(b) (black line), the $\Delta$PDF of the open structure at the CI is characterized by two enhanced peaks at approximately 2.8~\text{\AA} and 3.7~\text{\AA}. Due to the contributions from both the S$_2$ minimum and the open structure, the measured positive signal at 2.8~\text{\AA} has a higher intensity and longer duration than that at 1.9~\text{\AA}. The subsequent decrease of these signals can be attributed to dissociation as the molecule relaxes back to the ground state after passing through the CI. The signal at 3.7~\text{\AA}, associated with the open structure at the S$_1$/S$_0$ CI, appears to exhibit a delay with respect to that at 1.9~\text{\AA}, suggesting that the nuclear wave packet traverses from S$_2$ minimum to S$_1$/S$_0$ in approximately 100-150 fs.

\section{Branching ratio of the photoproducts}

In light of the inherent uncertainty associated with the conversion of the measured PD to $\Delta$PDF, the branching ratio of the various photoproducts is obtained through a global fitting of the PD signal to theoretical models. The measured $\Delta$PDF is exclusively utilized to guide the fitting process; that is to say, it dictates the selection of photoproducts to be employed in the fitting procedure. 

Figures~\ref{f6}(a), (b) and (c) illustrate the results at various time delays, with the data averaged over a 100~fs time window. As shown in Fig.~\ref{f5}(a) and Fig.~\ref{f5}(b), immediately following photoexcitation, the molecules are not yet dissociated. Consequently, in Fig.~\ref{f6}(a), the measured PD at $t\approx20~$fs is fitted with the combination of the theoretical PD for the S$_2$ minimum and the open structure at CI. The resultant PD from this fitting is in close agreement with the measurement, and the best fit indicates that the S$_2$ minimum structures account for approximately 75\% of the geometries of the molecules. 

A comparison of the measured PD at $t\approx130~$fs (Fig.~\ref{f6}(b)) with that at $t\approx20~$fs indicates a significant disparity in the ratio of the peak at $2~\text{\AA}^{-1}$ over that at $7.5~\text{\AA}^{-1}$. The increased ratio suggests that a greater number of molecules have undergone substantial structural alterations. As shown in Fig.~\ref{f5}(a) and Fig.~\ref{f5}(b), the majority of molecules have left the S$_2$ minimum, and a significant number have already undergone dissociation at $t\approx130~$fs. Therefore, the measured PD at $t\approx130~$fs is modeled using the combination of the theoretical PD for the photoproducts associated with C$_2$ and C$_3$ channels and the open structure at CI. The fitted results suggest that the majority of the molecules  ($63\pm16\%$) have not yet undergone dissociation at $t\approx130~$fs. The substantial uncertainty for the fitted results, particularly for the C$_2$ ($21\pm20\%$) and C$_3$ ($16\pm15\%$) channels, is presumably attributable to the theoretical PD assuming well-separated photoproducts for these channels. However, in the experiment with this short time delay, the photoproducts resulting from fragmentation are not significantly distant from each other. 

Following the results in Fig.~\ref{f4}(a) and Fig.~\ref{f5}(a), the measured PD at $t\approx1.1~$ps is modeled using the combination of the theoretical PD for the photoproducts associated with C$_2$ and C$_3$, as well as the ring-opened structures~\cite{peng2024}. The resultant PD from this fitting is in excellent agreement with the measurement, reproducing the further increased ratio of the peak at $2~\text{\AA}^{-1}$ over that at $7.5~\text{\AA}^{-1}$, and yielding a C$_3$:C$_2$:Ring-opened ratio of approximately 5:3:2. It should be noted that as time progresses, the ring-opened cyclobutanone may undergo further dissociation, thereby resulting in a change in the branching ratio. Consequently, the obtained branching ratio is only valid at this specific time delay.

\section{Conclusions and discussions}

In this study, we presented our experimental results on the photochemical dynamics of cyclobutanone upon excitation at 200 nm to the 3s Rydberg state using SJTU's MeV UED instrument. The objective of this work is to provide an independent measurement that, in conjunction with the independent experimental results from SLAC's MeV UED~\cite{green2025}, will validate the predictive capability of the simulations. It is found that the "Prediction Challenge" imposes a substantial challenge on experiments as well. This is due to the fact that the experimental results need to be accurate enough to serve as the "answer key" to assess the precision of theoretical predictions. In our experiment, the data within the range of $s<0.8~\text{\AA}^{-1}$ were not collected due to the hole in the center of the phosphor screen, which significantly impeded the ability to differentiate between ring-opening and dissociation. The primary distinction in the diffraction distribution is evident in the low-$s$ region, where the PD signal for ring-opening declines to zero while that for dissociation reaches a maximum as $s$ approaches zero. The inelastic signal, while enabling the tracking of the evolution of the electronic state, concomitantly increases the difficulty of obtaining a clean structure-change-related elastic scattering signal in the low-$s$ region, particularly during the early stages of the dynamics. The considerable uncertainty of the measured elastic scattering signal in the low-$s$ region may compromise the reliability of the PDF at a large distance. Consequently, as illustrated in Fig.~\ref{f4}(a) the $\Delta$PDF is displayed exclusively up to $4~\text{\AA}$. A further challenge for the experiment is to obtain the PD signal at the high-$s$ region with a sufficient signal-to-noise ratio to track the subtle structural changes at the early stage of the dynamics. The limited temporal resolution may also smear out the fine features in the PD and $\Delta$PDF, rendering an accurate comparison with theoretical predictions difficult. 

Having said that, we believe that our experiment has captured the primary dynamics of the photoinduced chemical reaction in cyclobutanone. The ensuing discussion will address the experimental findings that appear to be reliable and those that seem uncertain. We observed the inelastic scattering signal which encodes information about the electronic state. Despite the inability to quantify the individual contributions of S$_2$ and S$_1$ states to the inelastic signal in the absence of theoretical support, the falling edge of the signal suggests a sub-picosecond timescale for the photodissociation dynamics and an excited state lifetime of approximately 230~fs. While the existence of the inelastic signal is well-supported by the measured data, the obtained decay time may be subject to some uncertainty due to the contribution from elastic scattering, the removal of which may be model dependent. The dissociation through C$_2$ and C$_3$ channels is substantiated by the measured $\Delta$PDF where a positive signal at approximately $1.0~\text{\AA}$ and negative signals at approximately $1.6~\text{\AA}$ and $2.3~\text{\AA}$ are clearly observed. The presence of ring-opened cyclobutanone is also supported by the positive signal at $3.8~\text{\AA}$, although its intensity is only slightly  higher than the background. While there is confidence in the existence of these photoproducts, the branching ratio of C$_3$:C$_2$:Ring-opened obtained using the data averaged from 1.1~ps to 1.2~ps should be regarded as qualitative rather than quantitative, given that the C$_3$ channel comprises both cyclopropane and propene and their ratio may affect the model used in the fitting. Regarding the timescale of the dynamics, our data indicate that the dissociation process in the C$_2$ channel is completed much earlier than that in the C$_3$ channel. Furthermore, the observed modulation of the $\Delta$PDF around time zero and the delay of the positive signal at $r\approx3.8~\text{\AA}$ with respect to that at 1.9~\text{\AA} and  2.8~\text{\AA} suggests that we may have captured the nuclear wave packet traversing from S$_2$ minimum to S$_1$/S$_0$ CI. It is crucial to note that the experimental features in the $\Delta$PDF that support these interpretations are contingent on the PD signal in the high-$s$ regions, which is relatively weak in our measurement. In view of the considerable signal fluctuations observed around time zero resulting from the weak signal in the high-$s$ regions, it is necessary to perform further measurements to provide definitive evidence.  

Finally we wish to point out that the combination of inelastic signal and high-$s$ data from UED with the elastic scattering low-$s$ data from ultrafast x-ray scattering~\cite{ma2020ultrafast} may provide a comprehensive perspective on the photoinduced dynamics of cyclobutanone. It is important to note that while the theoretical predictions are conducted prior to the release of the experimental data, our data analysis is conducted after the theoretical predictions are available~\cite{eng2024,Vindel-Zandbergen2024,miao2024,peng2024,janos2024,hutton2024,makhov2024,suchan2024,jaiswal2024,lawrence2024,mukherjee2024,martin_santa2024,hait2024,miller2024,bennett2024}. We believe that a collaborative effort integrating theoretical and experimental methodologies will hold great potential for advancing the field of photophysics and photochemistry.

\begin{acknowledgments}
We thank J. Peng and Z. Lan for providing the simulated molecule structures and Z. He and G. Wu for assistance in optimization for the UV pump laser. This work is supported by the National Natural Science Foundation of China (Nos. 11925505, 12335010 and 11925405) and by the National Key R\&D Program of China (no. 2024YFA1612204). D.X. would like to acknowledge the support from the New Cornerstone Science Foundation through the Xplorer Prize. The UED experiment was supported by the Shanghai soft X-ray free-electron laser facility.  
\end{acknowledgments}

\section*{Data Availability Statement}

The raw data are stored at the MeV-UED facility of Shanghai Jiao Tong University.
All data supporting the conclusions are available from the corresponding author upon reasonable request.

\bibliographystyle{aipnum4-1}
\bibliography{refs}

\end{document}